# Traffic Analysis and Control at Proxy Server

Soumen Kanrar
Department of Computer Science
Vidyasagar University
Midnapour, India
rscs_soumen@mail.vidyasagar.ac.in

Niranjan Kumar Mandal
Department of Electrical Engineering
University of Engineering and Management
Calcutta, India
Niranjankumarmandal54@gmail.com

*Abstract*—Bandwidth optimization for real-time video traffic transmission through the proxy server is a challenging issue for the next- generation network. The self - control of the traffic rate at the proxy server is based on the relative data present in the proxy server or to import the require data from the remote data center node. The smoothness of traffic transmission is highly depended on the self -configuration in the cache memory at the proxy server by used of optimized page replacement procedure. The web proxy cache memory is finite in size. The system performance is highly depended upon the self- optimization of the traffic rate at the proxy server. This paper presents the effect of zipf- distribution parameters to the outbound request and the effect of the parameters to measure the bandwidth requirement for the desire video file. This paper presents some of the relevant results for bandwidth optimization and the effect of traffic control during active transmission for various sizes of video file through the proxy server.

## I. INTRODUCTION

The cost-effective system development for the smoothly video data stream transmission over the wired connection to maintain the grade of service depends on the load to the web cache server. The architecture of video on demand system used web cache to protect the collected data center nods to overcome various types of unwanted attack. The massive traffic load initially submitted to the web cache to fetch the desire video in full range or for the clips or taller/clips of any audio /video. Researchers concentrate over the huge growth of the Internet user and that bring enormous attention to explore the traffic control and bandwidth optimization for the real-time video streaming over the decades. A good number of works found in the area of network traffic handle. The hit ratio of a web cache grows to log-like fashion as a function on the cache size [1], [2], [3], [7], [8], [10], [13], [14], [15]. The equivalent capacity in the network and its application to bandwidth allocation in high speed networks is the grade of service [4], [16], [17]. The probability based admission control for the outbound traffic rate is used the tree based, and graph based network architecture [12]. The problem of bandwidth allocation in the high speed network has been addressed in literatures [4], [5], [6] in various ways by restrict to bounded ranges of the connection parameters. The true traffic control is done at the web proxy server for the limited buffer size of the cache memory. The paper is organized as brief description of the problem and literature survey in the introduction section I. Reaming section describes about the analytic description of the problem and the session initiative page replacement algorithm in section II. The section III presents the simulation results. The last section IV describes the about the conclusion remarks further improvement and reference.

## II. ANALYTIC DESCRIPTION OF THE MODEL

The web proxy traces for the random requests from the infinite heterogeneous population pull of client. Let $A$ be the given arrival of a page request and $B$ be the arrival request for page at web cache proxy server. The pages are ranked in order of their popularity (i.e. according to the hit count of the page). where page is the most popular page so the page are marked as $1, 1/2, 1/3, 1/4, \ldots, 1/(1+n), \ldots$ Now two mutually independent exclusive events occurs $A$ and $B$, $p(A) = \dfrac{1}{\sum_{i=1}^{N} \dfrac{1}{i}}$ and $p(B) = \dfrac{1}{i}$,

So, $p(AB) = \dfrac{1}{i} \cdot \dfrac{1}{\sum_{i=1}^{N} \dfrac{1}{i}}$ .

Here, $N$ is a finite natural number. So for the large size of population of submitted request and heterogeneous nature of the request pattern, that is better approximated according to reference paper [1]. Here,

$p(A) \approx \dfrac{1}{\sum_{i=1}^{N} \dfrac{1}{i^{\alpha}}}$ and $p(B) \approx \dfrac{1}{i^{\alpha}}$. Now,

$p_N(i) = p(AB) \approx \dfrac{1}{i^{\alpha}} \cdot \dfrac{1}{\sum_{i=1}^{N} \dfrac{1}{i^{\alpha}}}$ . This is the probability for the

requested page $i$ (say), $i \in I^{\succ 0}$, and assume that the finite stream of outbound requests $R$ is satisfied. The request $(R+1)^{th}$ is made only for the $i^{th}$ page, and it is considered that the page $i$ is not present in the cache memory of the web proxy server. So the distribution is obtained (or approximated) by the using the repetition of permutation $\approx (1 - p_N(i))^R \approx f(i)$ here $1 \leq R \leq N$, $f$ is a continuous real valued function.

The corresponding hit rate (miss) is expressed by the hit on demand basis as $H_{demand} = \sum_{i=1}^{N} p_N(i) f(i)$, Since the cache size is limited to $(C)$ say and the outbound request comes from the infinite size of heterogeneous population .i.e. $R \rightarrow \infty$, so hit miss be updated

As, $\lim_{R \rightarrow \infty} \sum_{i=1}^{C} p_N(i)\{1- p_N(i)\}^R$, Clearly $|1- p_N(i)| \prec 1$

The expanding series is convergent. So $H_{demand}$ is equivalent to $\sum_{i=1}^{C} p_N(i)$. If the required file is not present in the cache, two things are happened simultaneously, imported the file from data remote data center nodes connected to web proxy server and update the cache list the page. For the cache list update it is used by the dynamic session initiative page replacement algorithm. Figure -1 presents the work flow of the system in response to the outbound request.

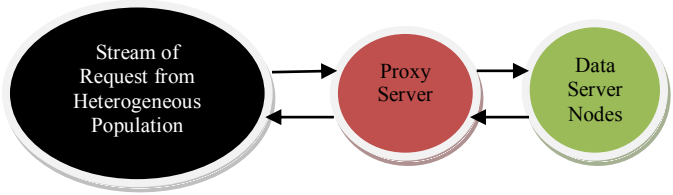

Fig. 1. *Traffic flows in the Video on Demand System*

So for the hit miss for $i^{th}$ file or page, considering $b_i$ is the required bandwidth to import the $i^{th}$ file from the remote data center node. Bandwidth $_{demand\,(i)} \approx b_i$, now, $b_i = s_i.t_i$ in ideal condition. Here, $S_i$ is the size of requested file and $t_i$ is the time duration of the activity of channel for transportation.

So, the aggregate bandwidth on demand is approximated by

$\approx \sum_{i=1}^{\infty} H_{demand(i)}.(s_i.t_i) = \sum_{i=1}^{\infty}\sum_{j=1}^{C} p_N(j)(s_i.t_i) = \sum_{i=1}^{\infty}\{\sum_{j=1}^{C} p_N(j)\}(s_i.t_i)$

$= k\sum_{i=1}^{\infty} \alpha.(C^{1-\alpha}).(s_i.t_i)$  (1)

$k$ is a threshold value related to packet loss, $0 \prec k \prec 1$ and $C$ is the finite cache memory size, $\alpha$ is the Zipf – parameter.

2.1 **Lemma**: For any real or complex value, $S$ is the expanding series $\zeta(s) = \sum_{n=1}^{\infty} n^{-s}$ satisfy $\zeta(s) \leq |s| n^{1-S}$

**Proof:** let $s$ be any complex variable then $s = \alpha + i\beta$,
$n^{-s} = \exp(\log\{n^{-s}\}) = \exp(-s \log n) = \exp\{(-\alpha - i\beta)\log n\}$
$= [\exp\{(-\alpha - i\beta)\}]^{\log n}$ by the rule $\exp(ab) = [\exp(a)]^b$
$= [\exp\{(-\alpha) + (-i\beta)\}]^{\log n}$

$= [\exp(-\alpha)]^{\log n} \; [\exp(-i\beta)]^{\log n} = [e^{-\alpha}]^{\log(n)}[e^{-i\beta}]^{\log(n)}$
$= [e^{\log(n)^{-\alpha}}] \, [e^{-i\beta}]^{\log n} = n^{-\alpha} \, e^{-i\beta \log n}$
$\Rightarrow n^{-s} = n^{-\alpha} e^{-i\beta \log n}$
So, $|n^{-s}| = |n^{-\alpha} e^{-i\beta \log n}| \leq n^{-\alpha} \|e^{-i\beta \log n}\| \leq n^{-\alpha}$

As, $\alpha > 0$ and $|e^{-i\beta \log n}| \rightarrow 1$

$\zeta(s)$ is analytic in the complex plane except a simple pole at $s = 1$. Now, $\zeta(s) \frac{1}{s-1} = \sum_{n=1}^{\infty}[n^{-s} - \int_{n}^{n+1} x^{-s} dx]$

$= \sum_{n=1}^{\infty} \int_{n}^{n+1} (n^{-s} - x^{-s}) dx$

for, $x \in [n, n+1]$, $(n \geq 1)$ and $\alpha \succ 0$

We get $|n^{-s} - x^{-s}| \leq |s \int_{x}^{n} y^{-1-s} dy| \leq |s| n^{1-\alpha}$

$|n^{-s} - x^{-s}| \leq |s \int_{x}^{n} y^{-1-s} dy| \leq |s| n^{1-\alpha}$

$\Rightarrow \zeta(s) - \frac{1}{s-1} \leq \sum_{n=1}^{\infty} |s| n^{1-\alpha}$

Only for real it is $\sum_{n=1}^{\infty} |\alpha| n^{1-\alpha}$

| Session Initiative Page Replacement Algorithm |
|---|
| Var:
List: array [1, 2.., n] of string
Hit_count: array [1, 2.., n] of real
Buffer: array [1, 2…, m] of string // m> n
Initialization:
List ← Store the name of the video file name
Begin
    While (buffer ← index! =Null)
    {
        Read the stream of request to buffer for a session
    }
    While (buffer ← index! = Null)
    {
        If (input string match with any member of the list)
{
Hit_count of the file ++
Stream transfer to the client
}
else
{
Step-1: Search the least hit count in Hit_count
Step-2: Replace with new file name entry

Step-3: Store the file to server
Step-4: Transfer the video to the client
}
Flash buffer
}
End |

## III. SIMULATION RESULT AND DISCUSSION MODEL

In this section, we describe the overall system response on the basis of different values of Zipf's parameters. The heterogeneous population size of the outbound request is $100^3$ for 10 milliseconds of time and the size as the file varies from (1 to 15 kb), the channel access time varies within (1 to 10 millisecond).

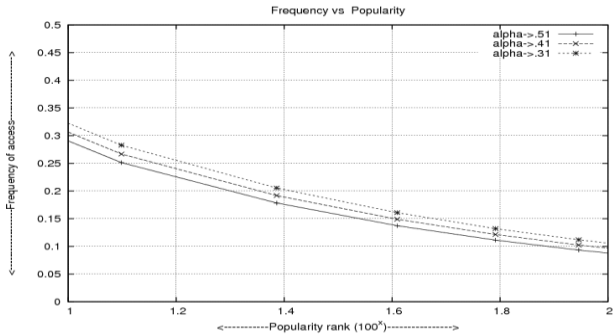

Fig. 2. Traffic Estimation against the Popularity low range

Figure 2 and 3 present the frequency of the user choice form the large scale of heterogeneous population of viewers. The traffic traces pattern done at the web proxy server. The Zipf parameter varies over the range (.98, .75, .64, .51, .41, .31) the proxy server has limited cache memory size. The popularity ranks in the scale $100^x$, at x=0, the top rank is 1. According to the plotted simulation figure 2 and figure 3 has a long tail that indicates the popularity decreases implies the frequency of accessing of the video decreases. The population size of the submitted request is $100^3$. The simulation shows that the hit misses increases when the value of $\alpha$ increases. The higher value of $\alpha$ indicates the hot spot situation in the proxy server for a particular session of the outbound request, where the submitted request for limited files is very high. The figure 2 and 3 indicates the file with rank (1 to $100^{1.1}$) and files with rank (1 to $100^{1.05}$) appears as hot spot situation. The request patterns are approximated by different values of $\alpha$. Figure 4 and figure 5 present the bandwidth requirement for the video transfer from the remote data center nodes to the heterogeneous client. The data center nodes are distributed, all the requested video streaming through a web cache proxy server.

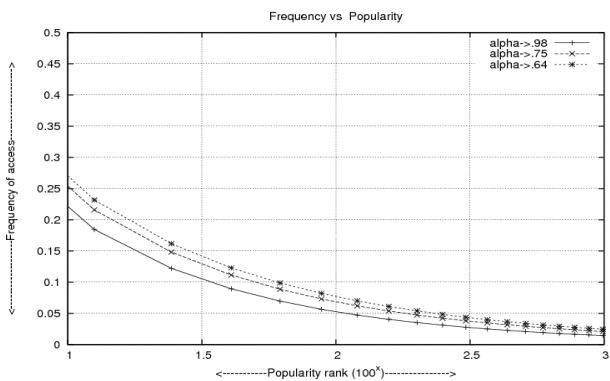

Fig. 3. Traffic Estimation against the Popularity low range

The simulation result in the figure 4 considers Zipf parameters as (.98 and .64). As usual for the figure 5, it is (.41 and .31). The high bandwidth consumes for the top 100 video files. For the simulation purpose, we consider the packet loss and inter packet delay threshold bounded within (0, 1) according to the expression (1). Figure 4 and figure 5 shows the comparative scenario for different values of $\alpha$.

Figure 4 indicates bandwidth requirement higher for the lower values of $\alpha$. The bandwidths are required for the file with the rank $100^{.475}$, 9.9 bits/sec for the $\alpha$ value .98 and 8.9 bits/sec for the $\alpha$ value .64. The simulation figure shows bandwidth requirement decrease according to the $\alpha$ value decrease. In the video on demand system major request submitted for the high rank video only. The major bandwidth required for the rank (1 to $100^{.75}$) video files.

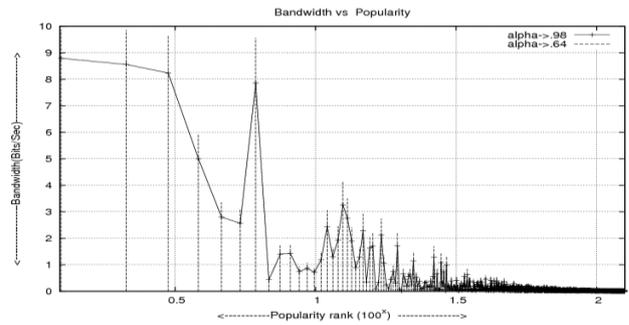

Fig. 4. Bandwidth Estimation against the Popularity high range

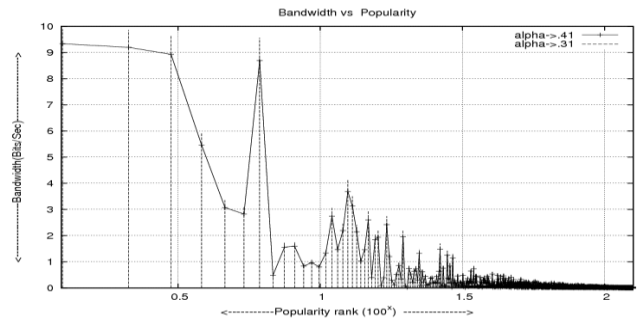

Fig. 5. Bandwidth Estimation against Popularity low range

The heterogeneous population size of the client is $100^3$ and the size as the file varies from (1 to 15 kb) the channel access time varies within (1 to 10 millisecond). Figure 4 and figure 5 has a long tail that indicated the bandwidth requirement coming less as the popularity goes decrease.

## IV. CONCLUSION REMARKS

In this paper, we have shown traffic request pattern submitted at the proxy server that matches with the analytic model of the system. We have considered the ideal situation for request response scenario. The hot spot space indicates for a limited number of files, and the required bandwidth is estimated for the hit miss and hit rate also. The delay and packet jitter estimation be added to the expression (1) in contrast with the real-world scenario for the further improvement of the paper. The distributed data center node and the routing of the packet to optimize the least hop count between the web proxy and the data center node will present the video streaming more realistic. The new types of algorithms needed to enhance the overall system performance.